\documentclass{epl}

\title{
$\sin{(2 \varphi)}$ current-phase relation in SFS junctions with
decoherence in the ferromagnet}
\shorttitle{$\sin{(2 \varphi)}$ current-phase relation in SFS junctions}

\author{R. M\'elin\thanks{regis.melin@grenoble.cnrs.fr}}

\institute{ 
Centre de Recherches sur les Tr\`es Basses
Temp\'eratures (CRTBT)\thanks{U.P.R. 5001 du CNRS, Laboratoire conventionn\'e
avec l'Universit\'e Joseph Fourier},\\ Bo\^{\i}te Postale 166, F-38042
Grenoble Cedex 9, France}


\begin{document}
\maketitle

\begin{abstract}
We propose a theoretical description of the $\sin{(2\varphi)}$
current-phase relation in SFS junctions at the $0$-$\pi$
cross-over obtained in recent experiments by Sellier
{\it et al.}~\cite{Sellier2} where it was suggested that
a strong decoherence in the magnetic alloy can explain the
magnitude of the residual supercurrent at the $0$-$\pi$
cross-over.
To describe the interplay between decoherence and
elastic scattering in the ferromagnet we use an analogy with
crossed Andreev reflection in the presence of disorder.
The supercurrent as a function of the length $R$ of the ferromagnet
decays exponentially over a length
$\xi$, larger than the elastic scattering
length $l_d$ in the absence of decoherence,
and smaller than the coherence length $l_\varphi$ in
the absence of elastic scattering on impurities.
The best fit leads to
$\xi \simeq \xi_h^{({\rm diff})}/3$, where $\xi_h^{({\rm diff})}$ is
exchange length of the diffusive system without decoherence
(also equal to $\xi$ in the absence of decoherence).
The fit of experiments works well for the amplitude of
both the $\sin{\varphi}$ and $\sin{(2 \varphi)}$ harmonics.
\end{abstract}

\section{Introduction}
When Cooper pairs from a superconductor (S) penetrate a ferromagnet (F)
at a SF interface the spin-up
electron decreases its spin energy because of Zeeman splitting
whereas the spin-down electron increases its spin energy. 
As a consequence the spin-up electron
accelerates and the spin-down electron decelerates so that
Cooper pairs acquire a finite center of mass momentum $\Delta k$.
This induces spatial oscillations of the pair amplitude
in the ferromagnet. If the length $R$ of the ferromagnet
is well chosen the Josephson relation of
a SFS junction can be inverted, giving
rise to a $\pi$-junction~\cite{Buzdin,Radovic,Heikkila,Golubov-short}.
The determination of the current-phase relation in SNS (where N
is a normal metal) and SFS junctions is a long standing problem
(see the recent review by Golubov {\it et al.}~\cite{Golubov-revue}).
The oscillations
of the transition temperature of FS superlattices~\cite{osc} as a 
function of the thickness of the ferromagnet are another
consequence of the $\pi$-coupling.

SFS $\pi$-junction have been probed recently in experiments
in which the ferromagnet is a magnetic alloy with a sufficiently
small exchange field~\cite{Aarts,Rya,Kontos,Guichard,Sellier1}
so that the period of the spatial oscillations of the pair amplitude
is large enough.
In the following we discuss a recent
experiment~\cite{Sellier2} in which half-integer Shapiro
steps were observed in a Nb/CuNi/Nb junction at
the $0$-$\pi$ cross-over,
indicating a $\sin{(2 \varphi)}$ current-phase relation.
For highly transparent interfaces and without decoherence
the current-phase relation is related to 
the derivative of the energy of the
Andreev bound states with respect to
the superconducting phase difference between the two
superconductors~\cite{Radovic,Heikkila,Golubov-short,Golubov-revue}.
The exchange field in the ferromagnet
generates a Zeeman splitting of the spectrum of Andreev
bound states, which was the explanation of the 
$\sin{(2\varphi)}$ harmonics at the $0$-$\pi$ cross-over
proposed by Sellier {\sl et al.}~\cite{Sellier2}.
To explain the tiny magnitude of the residual supercurrent
at the 0-$\pi$ cross-over, Sellier {\sl et al.}~\cite{Sellier2}
suggested
the existence of a strong decoherence in the magnetic
alloy. However within this assumption the
Andreev bound states
are broadened so that the supercurrent cannot be anymore
evaluated as the derivative of the bound state energies.
This calls for a specific modeling of $\pi$ junction involving
decoherence in the ferromagnet.

The origin of
decoherence in a magnetic alloy is not well characterized
experimentally, because it is not possible to carry out
weak localization and universal conductance fluctuations
as a function of an applied magnetic field. Quantum coherence
in a ferromagnet was however studied in a recent experiment using
time-dependent universal conductance fluctuations~\cite{Lee}.
One source of decoherence in a ferromagnet is spin waves but
other effects can play a role such as
spatial heterogeneities of the exchange field or domain wall
motion~\cite{Hong,Tatara}. 

We discuss here the effect
of decoherence in the magnetic alloy, motivated by the
experimental observations of a very small residual
supercurrent at the
$0$-$\pi$ cross-over and more generally to a huge reduction of the
supercurrent compared to a model not involving
decoherence~\cite{Sellier2,Sellier1}.
We show that decoherence in the ferromagnet
implies the existence of a length $\xi$,
intermediate between (i) the elastic mean free path $l_d$ due to
elastic scattering on impurities in the absence of decoherence,
and (ii) the coherence length $l_\varphi$ due to decoherence
in the magnetic alloy in the absence of elastic scattering
on impurities.
The $\sin{\varphi}$ harmonics due to the coherent transfer of
a charge-$(2e)$ is proportional to 
$\exp{(-R/\xi)}$, and the $\sin{(2 \varphi)}$ harmonics
due to the coherence transfer of a charge-$(4e)$
is proportional
to $\exp{(-2 R / \xi)}$, where $R$ is the length of the ferromagnet.
The length scale $\xi$ can be smaller than $R$ even though
$l_\varphi$ is larger than $R$. 
Therefore we base our description on the first two terms of an
expansion in the coherent
transfer of multiples of a charge-$(2 e)$.
Similarly to Ref.~\cite{Melin-Feinberg-PRB} the dressing by 
multiple local Andreev reflections is described non perturbatively
so that our description is valid for highly transparent interfaces.
We also treat rigorously the geometrical effects
related to propagation parallel to the
interfaces.

\begin{figure}
\onefigure[width=11cm]{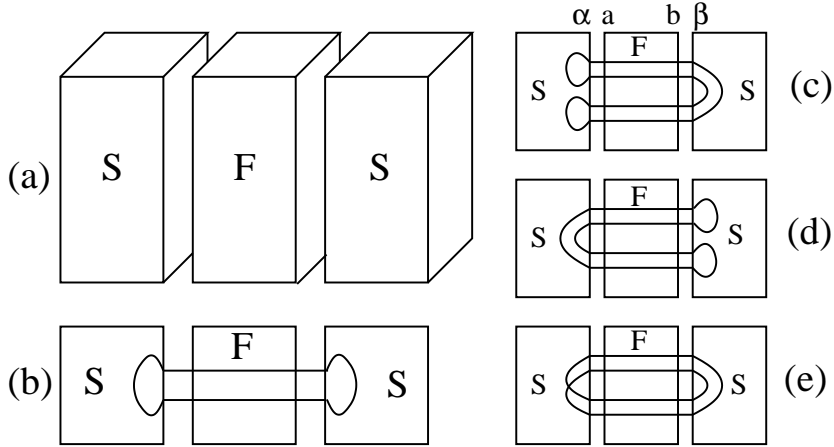}
\caption{Schematic representation of the 3D SFS junction (a);
the diagram contributing to the $\sin{\varphi}$ harmonics (b); 
the three diagrams contributing to the $\sin{2 \varphi}$
harmonics (c), (d), (e). 
For clarity, renormalization by local process is not shown on these
diagrams. The sites at the interfaces are labeled by
$\alpha_i$ and $a_i$ (interface $a$-$\alpha$) and
$b_i$, $\beta_i$ (interface $b$-$\beta$).
\label{fig:schema}
}
\end{figure}

\section{The models}

We consider a model in which a 3D ferromagnet is connected to two 3D
superconductors (see Fig.~\ref{fig:schema}-(a)).
The superconducting electrodes are described
by the BCS Hamiltonian
\begin{equation}
{\cal H}_{\rm BCS} = \sum_{\langle \alpha , \beta
\rangle , \sigma} - t \left(
c_{\alpha,\sigma}^+ c_{\beta,\sigma}
+c_{\beta,\sigma}^+ c_{\alpha,\sigma} \right)
+ \Delta \sum_\alpha \left(
c_{\alpha,\uparrow}^+ c_{\alpha,\downarrow}^+
+ c_{\alpha,\downarrow}
c_{\alpha,\uparrow} \right)
,
\end{equation}
where $\alpha$ and $\beta$ correspond to neighboring sites on a
cubic lattice.
The ferromagnetic electrode is described by the Stoner model
\begin{equation}
\label{eq:Stoner}
{\cal H}_{\rm Stoner} = \sum_{\langle \alpha,\beta
\rangle,\sigma} -t \left( c_{\alpha,\sigma}^+
c_{\beta,\sigma} + c_{\beta,\sigma}^+ 
c_{\alpha,\sigma} \right)
-h_{\rm ex} \sum_\alpha
\left( c_{\alpha,\uparrow}^+ c_{\alpha,\uparrow}
-c_{\alpha,\downarrow}^+ c_{\alpha,\downarrow} \right)
,
\end{equation}
where $h_{\rm ex}$ is the exchange field.
We add to (\ref{eq:Stoner}) a term describing elastic
scattering by impurities.
The ferromagnet is connected
to the superconducting layers by the Hamiltonian
\begin{equation}
\label{eq:tunnel-ab}
{\cal W}_{a (b)} = \sum_{\alpha,\sigma} - t_{a (b)} \left(
c_{\alpha,\sigma,a (b)}^+ c_{\alpha,\sigma,S} +
c_{\alpha,\sigma,S}^+ c_{\alpha,\sigma,a (b)} \right) 
,
\end{equation}
where the label $\alpha$ runs over all sites at the interface.

To describe propagation in the ferromagnet we use
the advanced Green's function of a
bulk ballistic 3D ferromagnet
\begin{equation}
\label{eq:g-3D}
g_{a,b}^{1,1,A} = -\frac{1}{t_F} \frac{1}{k_F^\uparrow R}
e^{-i k_F^\uparrow R} e^{-R/l_\varphi}
,
\end{equation}
where $l_\varphi$ is the phase coherence length.
A similar expression is obtained for $g_{a,b}^{2,2,A}$.
We will discuss how to incorporate disorder in 
the ferromagnetic electrode. The Green's functions of a superconductor
can be found in the literature~\cite{Abrikosov}.

The Nambu representation of the tunnel amplitudes is given by
\begin{equation}
\hat{t}_{a,\alpha} = \hat{t}_{\alpha,a}^*=t_a \left(
\begin{array}{cc} e^{i \varphi /4} & 0 \\
0 & -e^{-i \varphi / 4} \end{array} \right),
\end{equation}
and similar equations are obtained for $\hat{t}_{b,\beta}$
and $\hat{t}_{\beta,b}$.

\section{Supercurrent}
The supercurrent is given by
\begin{equation}
\label{eq:sp-I-def}
I_S=\frac{e}{h} \int_0^{+\infty}
\mbox{Tr}
\left\{ \hat{\sigma}^z \left[ 
\hat{t}_{\alpha,a} \left( \hat{G}^A_{a,\alpha} 
-\hat{G}^R_{a,\alpha} \right) 
- \hat{t}_{a,\alpha} \left( \hat{G}^A_{\alpha,a} 
-\hat{G}^R_{\alpha,a} \right) \right] \right\}
d \omega
\nonumber
,
\end{equation}
where the trace corresponds to
a summation over the ``11'' and
``22'' matrix elements in the Nambu representation, a 
summation over the two spin orientations (formally equivalent to
summing over the ``33'' and ``44'' components in the spin 
$\otimes$ Nambu representation~\cite{Melin-Peysson}),
and a summation over the channel labels.
Eq.~(\ref{eq:sp-I-def}) can be derived from
Keldysh formalism~\cite{Cuevas,Nozieres}
by noting that in equilibrium the Keldysh Green's function
takes the simple form
$\hat{G}^{+,-}=n_F(\omega) \left[
\hat{G}^A-\hat{G}^R \right]$.
The fully dressed Green's functions $\hat{G}_{i,j}$
are determined through the Dyson equation $\hat{G}=\hat{g}+\hat{g}
\otimes \hat{\Sigma} \otimes \hat{G}$, 
where $\hat{\Sigma}$ is the self energy corresponding to the
bonds $t_{a,\alpha}$ and $t_{b,\beta}$ and $\otimes$ denotes a
summation over the sites $\alpha_i$, $a_i$, $b_i$, $\beta_i$ 
(see Fig.~\ref{fig:schema}) and
a convolution over the time arguments that becomes a simple product
once the Dyson equation is Fourier transformed to 
the energy variable $\omega$. The supercurrent takes the simpler form
\begin{equation}
I_S=\frac{e}{h} \int_0^{+\infty}
\mbox{Tr}\left\{ \hat{t}_{a,\alpha}
[ \hat{g}_{\alpha,\alpha}^A, \hat{\sigma}^z ]
\hat{t}_{\alpha,a} \hat{G}_{a,a}^A
- \hat{t}_{a,\alpha}
[ \hat{g}_{\alpha,\alpha}^R, \hat{\sigma}^z ]
\hat{t}_{\alpha,a} \hat{G}_{a,a}^R \right\}
,
\end{equation}
where $[]$ is an anticommutator and $\hat{\sigma}^z$
one of the Pauli matrices.

\section{Perturbative expansion of the supercurrent}
Now we describe a perturbative expansion in which we include
the coherent transfer of a charge-$(2e)$ and charge-$(4e)$ object
while keeping a non perturbative description of local processes.
The Green's function $\hat{G}_{a,a}$ can be expanded according to
$\hat{G}_{a,a}=\sum_n \hat{G}_{a,a}^{(n)}$, where 
$\hat{G}_{a,a}^{(n)}$ describes the $\sin{(n \varphi)}$ harmonics
due to the coherent transfer of a charge-$(2ne)$ object.
We obtain $\hat{G}_{a,a}^{(0)}=\hat{K}_{a,a} \hat{g}_{a,a}$,
\begin{eqnarray}
\hat{G}_{a,a}^{(1)}&=&
\hat{K}_{a,a} \hat{X}_{a,b} \hat{K}_{b,b}
\hat{X}_{b,a} \left[ \hat{g}_{b,a}
+\hat{X}_{b,a} \hat{K}_{a,a} \hat{g}_{a,a} \right]\\
\hat{G}_{a,a}^{(2)}&=& 
 \hat{K}_{a,a} \hat{X}_{a,b} \hat{K}_{b,b}
\hat{X}_{b,a} \hat{K}_{a,a} \hat{X}_{a,b} \hat{K}_{b,b}
\left[ \hat{g}_{b,a}
+ \hat{X}_{b,a} \hat{K}_{a,a} \hat{g}_{a,a} \right]
,
\end{eqnarray}
with $\hat{X}_{a,a}=\hat{g}_{a,a} \hat{t}_{a,\alpha}
\hat{g}_{\alpha,\alpha} \hat{t}_{\alpha,a}$,
$\hat{K}_{a,a}=\left[\hat{I}-\hat{X}_{a,a}\right]^{-1}$, and with
similar expressions for $\hat{K}_{b,b}$, $\hat{K}_{a,b}$,
$\hat{K}_{b,a}$. The channels labels are kept implicit.

The $\sin{\varphi}$ harmonics of the supercurrent is given by
$I_S^{(1)} \sin{\varphi}$, with
\begin{equation}
I_S^{(1)}={e \over h} \int_0^{+\infty} d \omega
\sum_{a,b}
\left\{
\mbox{Tr} \left[ \hat{g}_{a,b} \hat{V}_{b,b}
\hat{g}_{b,a} \hat{W}_{a,a} \right]
+\mbox{Tr} \left[ \hat{g}_{a,b} \hat{V}_{b,b}
\hat{g}_{b,a} \hat{W}'_{a,a} \right] \right\}
,
\end{equation}
where the sum over $a$ and $b$ is a sum over all channels
at the two interfaces.
The vertices $\hat{V}$, $\hat{W}$ and 
$\hat{W}'$ containing information about the dressing
by local processes are given by
\begin{eqnarray}
\hat{V}_{a,a} &=& \hat{t}_{a,\alpha}
\hat{g}_{\alpha,\alpha} \hat{t}_{\alpha,a}
\hat{K}_{a,a}\\
\hat{W}_{a,a} &=& \hat{t}_{a,\alpha}
\left[\hat{g}_{\alpha,\alpha},\hat{\sigma}^z\right]
\hat{t}_{\alpha,a} \hat{K}_{a,a}\\
\hat{W}'_{a,a} &=& \hat{t}_{a,\alpha}
\hat{g}_{\alpha,\alpha} \hat{t}_{\alpha,a}
\hat{K}_{a,a} \hat{g}_{a,a} \hat{t}_{a,\alpha}
\left[\hat{g}_{\alpha,\alpha},\hat{\sigma}^z\right]
\hat{t}_{\alpha,a} \hat{K}_{a,a}
.
\end{eqnarray}
The same perturbative expansion can be applied to the amplitude
$I_S^{(2)}$ of the $\sin{(2\varphi)}$ harmonics.
The three diagrams on Fig.~\ref{fig:schema}-(c), (d), (e)
give rise to 24 terms that can all be evaluated explicitely.
One of these terms is
\begin{equation}
{e \over h} \int_0^{+\infty} d \omega
\sum_{a_1,b_1,a_2,b_2}
g_{a_1,b_1}^{1,1} V_{b_1,b_1}^{1,2}
g_{b_1,a_1}^{2,2} V_{a_1,a_2}^{2,1}
g_{a_2,b_2}^{1,1} V_{b_2,b_2}^{1,2}
g_{b_2,a_2}^{2,2} W_{a_2,a_1}^{2,1}
.
\end{equation}
The quantities $V_{b_1,b_1}^{1,2}$ and
$\int V^{2,1}(R) W^{2,1}(R) d {\bf R}$
(with ${\bf R}$ the vector between $a_1$
and $a_2$) are evaluated by a Fourier transform.
$V_{b_1,b_1}^{1,2}$ is proportional to
$\int V^{1,2}(k_\parallel) d {\bf k}_\parallel$ and
$\int V^{2,1}(R) W^{2,1}(R)d {\bf R}$ is proportional to
$V^{2,1}(k_\parallel=0) W^{2,1}(k_\parallel=0)$.
We now evaluate the average over disorder of
$\overline{g_{a_1,b_1}^{1,1} g_{b_1,a_1}^{2,2}}$
and show that, like for crossed Andreev
reflection~\cite{Feinberg,Cht},
disorder changes the value of the exponent in the
geometrical prefactor compared to the ballistic case    
and therefore enhances the supercurrent through the ferromagnet
compared to the ballistic case. We also calculate the coherence
length $\xi$ and the effective Fermi wave-vector mismatch
$\Delta K$ in the presence of both disorder and decoherence.

\section{Effect of disorder in the ferromagnet}
The average Green's function as a function of separation $d$
decays like $\overline{g_{a,b}^{1,1,A}}(d)=
g_{a,b}^{1,1,A}(d) \exp{(-d/l_d)}$~\cite{Abrikosov} where
$g_{a,b}^{1,1,A}(d)$ is given by Eq.~(\ref{eq:g-3D}).
The mean free path $l_d$ due to elastic scattering
on impurities is supposed to be much smaller
than the ballistic coherence length $l_\varphi$. 
The calculation of the diffusion probability
\begin{equation}
{\cal P}(d)=\frac{t_F}{a_0^3}
\overline{g_{1,1}^A(d) g_{2,2}^A(d)}
,
\end{equation}
with $a_0$ the lattice spacing,
is formally
analogous to the calculation of the subgap conductance of a
disordered
superconductor~\cite{Feinberg,Cht},
except for different phase factors. Following Ref.~\cite{Feinberg}
we obtain
\begin{equation}
{\cal P}(d) = - \frac{1}{4 \pi \hbar D d} \exp{\left(-\frac{d}
{\xi}\right)}
,
\end{equation}
with $D=v_F l_d / 3$ the diffusion constant, and with
\begin{equation}
\label{eq:xi}
\frac{1}{\xi} = \sqrt{\frac{3}{l_d} \left[
\frac{2}{l_\varphi}+i \Delta k \right]}
.
\end{equation}
We deduce from Eq.~(\ref{eq:xi}) the following expression of
the effective wave-vector mismatch $\Delta K$ and
coherence length $\xi$:
\begin{eqnarray}
\Delta K &=& \sqrt{\frac{3}{2 l_d}}
\sqrt{ \sqrt{ \left(\frac{2}{l_\varphi}\right)^2
+\left(\Delta k
\right)^2} - \frac{2}{l_\varphi}}\\
\frac{1}{\xi} &=&
\sqrt{\frac{3}{2 l_d}} 
\sqrt{ \sqrt{ \left(\frac{2}{l_\varphi}\right)^2
+\left(\Delta k \right)^2} + \frac{2}{l_\varphi}}
.
\end{eqnarray}
Integrating over all channels at the two interfaces we obtain
\begin{eqnarray}
\sum_{a,b} \overline{g_{a,b}^{1,1,A}(d) g_{b,a}^{2,2,A}(d)}
&=& - N_{\rm ch} \frac{a_0}{t_F} \frac{1}{4 \pi \hbar D}
\int \frac{2 \pi y dy}{\sqrt{R^2+y^2}}
\exp{(-d/\xi)} \exp{(i \Delta K d )}\\
&=& - N_{\rm ch} \frac{a_0}{t_F} \frac{1}{2 \hbar D}
\frac{\exp{(-R/\xi)} \exp{(i \Delta K R )}}{1/\xi-i \Delta K}
,
\end{eqnarray}
with $d=\sqrt{R^2+y^2}$.

\section{Fit of the experiments}
\begin{figure}
\onefigure[width=13cm]{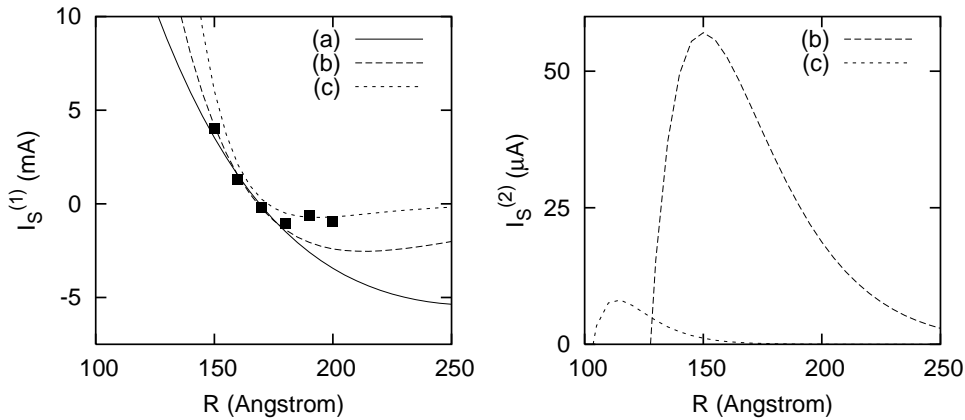}
\caption{Variation of the amplitudes $I_S^{(1)}$
and $I_S^{(2)}$ of the
$\sin{\varphi}$ and $\sin{(2\varphi)}$
harmonics 
as a function of the thickness $R$ of the ferromagnet.
The experimental points~\cite{Sellier1,Sellier2} are indicated
on the left panel. We use $D=3.5$~cm$^2$s$^{-1}$, $l_d=10$~\AA.
(a) corresponds to $h_{\rm ex}=70$~K,
$l_\varphi=7150$~\AA, $\xi=100$~\AA. 
(b) corresponds to $h_{\rm ex}=190$~K,
$l_\varphi=1300$~\AA, $\xi=45$~\AA.
(c) corresponds to $h_{\rm ex}=460$~K,
$l_\varphi=270$~\AA, $\xi=20$~\AA.
\label{fig:fit}
}
\end{figure}

Now we use the model discussed above to fit
quantitatively the recent experiments
by Sellier {\it et al.}~\cite{Sellier2}.
The value of the bulk hopping amplitudes $t_F$ and $t_S$ are chosen
equal to avoid multiplying the number of parameters, and such
that the diffusion constant of the ferromagnet is close to
$D=4$~cm$^2$s$^{-1}$~\cite{Sellier2}.
We use $t_S=t_F=5 \times 10^5$~K and
the elastic mean free path is $l_d=10$~\AA~\cite{Sellier1}.
The Fermi wave-vector
$k_F$ in the superconductor is chosen equal to $k_F=1$~\AA$^{-1}$,
and the lattice parameter is chosen
equal to $a_0=1$~\AA.
Since the interfaces
are highly transparent we choose $t_a=t_b=t_S=t_F$.
We fix the ratio between $l_\varphi$ and
$\xi_h^{({\rm ball})}=\hbar v_F/h_{\rm ex}$
in the ballistic system
to be smaller than unity,
and determine the exchange field in such a way that
the $\sin{\varphi}$ harmonics vanishes for the same value of $R$
as in the experiment.
We first tried a fit with $l_\varphi=\xi_h^{({\rm ball})}=7150$~\AA, 
$h_{\rm ex}=70$~K, and with
$\xi=100$~\AA. 
This fit is not compatible with the three experimental
points with the largest values of $R$ and the residual value of the
supercurrent at the $0$-$\pi$ cross-over is far too large (curve (a)
on Fig.~\ref{fig:fit}). To obtain a better agreement with experiments
we increase $h_{\rm ex}$ and decrease $l_\varphi$. The fit (b)
corresponds to $h_{\rm ex}=190$~K, $l_\varphi=1300$~\AA,
$\xi=45$~\AA. The fits (c) on Fig.~\ref{fig:fit} is even better, with
$h_{\rm ex}=460$~K, $l_\varphi=270$~\AA,
$\xi=20$~\AA.
The values of $\xi$ can be compared to the values of
$\xi_h^{({\rm diff})}=\sqrt{\hbar D / h_{\rm ex}}$
for the diffusive system in the absence of decoherence.
The values of $\xi_h^{({\rm diff})}/\xi$ are approximately equal
to $1.5$ (fit (a)), $2$ (fit (b)) and $3$ (fit (c)), whereas
$\xi_h^{({\rm diff})}/\xi=1$ in the absence of decoherence.

The amplitude of the $\sin{(2 \varphi)}$ harmonics at the 
$0$-$\pi$ cross-over is $4$~$\mu$A in experiments.
The fits (a) and (b) lead to much larger values whereas the fit
(c) has the correct order of magnitude (see Fig.~\ref{fig:fit}).
The agreement of the fits
of the amplitudes of both the $\sin{\varphi}$ and
$\sin{(2 \varphi)}$ harmonics suggest the
validity of the fit (c). The renormalization of the coherence length
$\xi$ compared to its value $\xi_h^{({\rm diff})}$ in the absence
of decoherence may depend on the quality of the magnetic layer
since it is expected qualitatively that decoherence is reduced
with less inhomogeneities
in the exchange field.

\section{Conclusion}
To conclude we have provided a modeling of the
$\sin{(2 \varphi)}$ current-phase relation in SFS junctions
at the $0$-$\pi$ cross-over. This model is motivated by the
fact that the residual supercurrent at the 0-$\pi$ cross-over
is very small, therefore suggesting that decoherence in the
ferromagnet plays a relevant role, as suggested
by Sellier {\sl et al}~\cite{Sellier2}.
Decoherence was introduced  
through a phenomenological coherence length $l_\varphi$.
Elucidating precisely its microscopic origin is difficult
experimentally but several factors (spin waves, inhomogeneities in
the exchange field, motion of domain walls) may play a role.
The supercurrent was calculated through
an expansion in the number of non local processes connecting 
the two interfaces. Lowest order processes correspond to a 
coherent transfer of a charge-$(2 e)$ contributing to the
$\sin{\varphi}$ harmonics and the next order corresponds
to a coherent transfer of a charge-$(4 e)$ contributing to the
$\sin{(2 \varphi)}$ harmonics. Like for crossed Andreev
reflection~\cite{Melin-Feinberg-PRB} these non local processes
are dressed by multiple local Andreev reflections that were
included in a non perturbative fashion. We also included
the geometrical
effect of propagation parallel to the interfaces.

We found that in the ferromagnet disorder effects (characterized
by the elastic scattering length $l_d$ in the absence of decoherence)
and decoherence 
(characterized by the coherence length $l_\varphi$ in the ballistic
limit) generate
a new coherence length $\xi$, intermediate between
$l_d$ and $l_\varphi$, and smaller than the exchange length
$\xi_h^{({\rm diff})}$
evaluated in the diffusive limit in the absence of decoherence.
The amplitude of the $\sin{\varphi}$ harmonics is
proportional to $\exp{(-R/\xi)}$ and the
amplitude of the $\sin{(2 \varphi)}$ harmonics is
proportional to $\exp{(-2 R/\xi)}$ so that
$\exp{(-R/\xi)}$ is a small parameter in the expansion
in non local Andreev processes. The approach is consistent
in the sense that the fit to experiments shows that
$\exp{(-R/\xi)}$ is a very small parameter [we have 
$R/\xi=8.75$ for the fit (c) on Fig.~\ref{fig:fit}] so that
the $\sin{(3 \varphi)}$ and higher harmonics can indeed
be neglected.

\stars

The author acknowledges fruitful discussions with
C. Baraduc, D. Feinberg, and H. Sellier
and thanks H. Courtois for a critical reading of the
manuscript.

\end{document}